\documentclass[12pt,preprint]{aastex}

\newcommand{\kms}{km s$^{-1}$}

\slugcomment{Revised version submitted to ApJL.}

\shorttitle{Runaway Stars in the Orion Nebula Cluster.}
\shortauthors{O'Dell, et al.}

\begin{document}


\title{HST Images Do Not Support the Presence of Three High Velocity, Low-Mass Runaway Stars in the Core of the Orion Nebula Cluster
\footnote{
Based in part on observations with the NASA/ESA Hubble Space Telescope,
obtained at the Space Telescope Science Institute, which is operated by
the Association of Universities for Research in Astronomy, Inc., under
NASA Contract No. NAS 5-26555.}}


\author{C. R. O'Dell}
\affil{Department of Physics and Astronomy, Box 1807-B, Vanderbilt University, Nashville, TN 37235}
\email{cr.odell@vanderbilt.edu}

\author{Arcadio Poveda and Christine Allen}
\affil{Instituto de Astronom\'{\i}a, UNAM, Apdo. Postal 70-264, 04510 Mex\'{\i}co DF, Mex\'{\i}co}

\and
\author{Massimo Robberto}
\affil{Space Telescope Science Institute, Baltimore, MD 21218}


\begin{abstract}

A recent article has employed the determination from groundbased images of high proper motions in the 
Orion Nebula Cluster to argue that JW 349, JW 355, and JW 451 are high velocity (38 \kms, 89 \kms, and 69 \kms, respectively) low mass 
runaway stars. We report on measurement of the proper motions of these stars using images made by the 
Hubble Space Telescope's WFPC2 imager and find that there is no evidence for motions above 6.2 \kms\ for JW 349 and 7.9 \kms\ for JW 355, while the motion of 5.5 \kms\ for JW 451 is only slightly larger
than the measurement uncertainty of 3.9 \kms. We conclude that there is no observational support 
for these stars being high velocity runaway stars. 

\end{abstract}


\keywords{astrometry---stars:kinematics---stars:pre-main-sequence}



\section{Introduction}

A recent study of proper motions of stars near the center of the Orion Nebula
Cluster (ONC, sometimes referred to as the Trapezium Cluster) revealed the fact
that three low mass stars appear to be moving at high tangential velocities \citep{pov05}.
These high velocities naturally raise the possibility that these are runaway stars
that have obtained their velocities through a dynamic process involving close young 
stars \citep{pov67}.
Recent studies have firmly established \citep{rod05,gom05} the presence of high velocity 
objects in the BN-KL grouping of infrared stars that is imbedded to the north-west
from the center of the ONC, but successfully disputed the arguments for these objects 
presented by \citep{tan04}. 

The Poveda study (Poveda et~al. 2005, henceforth PAH-A)
used the results of the photographic astrometry of Burton Jones and Merle Walker
(Jones \&\ Walker, 1988; henceforth JW). JW used 47 images made at the prime focus
of the Lick Shane 3 m reflector over the intervals 1960.1-1961.9 and 1981.8-1983.0.
Because of the high galactic latitude of the cluster (b$\approxeq$-20\arcdeg) and the fact that
it is located on the near side of a molecular cloud that is optically thick even
at the near infrared wavelengths (103aU or IN emulsions) used for the imaging, there are relatively
few field stars contaminating the sample and most of the stars lie within 
a proper motion vector of 0.2\arcsec/century. Stars with motions lying far outside
of this value were classified as field-stars, rather than cluster members. It was
only appropriate that PAH-A would re-assess the astrometric results
with the intention of seeing if any of the high proper motion stars were actually
cluster members with high spatial velocities. They identified three candidate stars
(JW 349, JW 355, and JW 451) for being runaways. These objects not only have large
proper motions in JW, but must be members of the ONC because the presence of 
associated ionized gas indicates that they are proplyds, ionized by the O6 star 
in the Trapezium $\Theta^{1}$ Ori~C \citep{ode94}.  

The importance of this result, that there are objects in the ONC identified as 
runaway stars, is such that it seemed appropriate to examine the observational arguments
for this conclusion. Fortunately the WFPC2 \citep{hol95} of the Hubble Space Telescope (HST) 
has been used frequently for imaging the ONC since soon after its installation in 
December, 1993 so that the time base of observations for astrometry can be about 
ten years. This is only about half of the time base for the JW observations, but,
the resolution of the WFPC2 images is more than ten times better than characteristic
of the Shane telescope without seeing-compensation. Not only does the better 
resolution allow a more accurate position determination, but it also allows avoiding
confusion with other objects in the vicinity, such as structure in the nebula.
This means that the WFPC2 observations potentially offer valuable information that 
allows testing the reliability of the reported high proper motion stars.

\section{Observations}
In this study, we draw upon two sets of observations, the first clustered soon after
installation of the WFPC2 and the second being quite recent, giving us the widest
possible time base for comparison of first and second epoch images.
Three early image sets were made under programs 
GO 5085 (Principal Investigator C. R. O'Dell),  GO 5193 (an Early Release Observations program), and GO 5469 (Principal Investigator John Bally).
Two recent programs have re-imaged the same areas near the Trapezium (GO 9141, Principal Investigator 
C. R. O'Dell; GO 10246, Principal Investigator Massimo Robberto).  

\subsection{First Epoch Observations}
The first epoch field for JW 349 was taken from the GO 5085 (pointing 5, \citep{ode96}) F656N image in CCD2. This was composed from two
200 second images combined to produce good cosmic ray event correction. They were made on 1995 January 19.
The first epoch field for JW 355 was taken from the GO 5469 (LV3 pointing, \citep{ode96}) F547M image in CCD2.
This was composed from three 30 second images combined to produce good cosmic ray event correction. They were made
on 1995 March 21.
The first epoch field for JW 451 was taken from the GO 5193 F547M image in CCD4. This was composed from two 100 s images combined to 
produce good cosmic ray event correction. They were made on 1993 December 29.

\subsection{Second Epoch Observations}
The second epoch observations used in our analysis were drawn from two recent
programs. Program GO 9141 \citep{ode03} imaged a single field centered south-west of the Trapezium. 
The much larger program GO 10246 \citep{rob04} has recently imaged over 100 fields that cover
the entire inner region of the Orion Nebula Cluster with both the WFPC2 and ACS and partially cover 
this same field with NICMOS.
The second epoch field for JW 349 was the WFPC2 F656N image in CCD4 of program GO 9141, this image
having been made at the same pointing with a pair of 140 second exposures on 2002 January 22. 
The second epoch
field for JW 355 used the WFPC2 F336W, F439W, and F814W images in CCD3 of program GO 10246 (position 38), combined to produce good cosmic ray event correction and increase the signal as the exposures were 400 s (twice), 80 s, and 10 s, respectively. These were made on 2005 April 10. 
The same field was used for the second epoch image of JW 451, except that it was selected from CCD2.


%% This section also includes several of the displayed math environments
%% mentioned in the Author Guide.

\section{Determination of Proper Motions of the Three Targeted Stars}

The WFPC2 images are quite stable within a single CCD detector, but the relative position
of the CCD's on the plane of the sky drift with time. This means that one cannot
use the entire mosiac for alignment of first and second epoch images. We have,
therefore, adopted a procedure used previously for measurement of shocks and jets
\citep{bal00,ode02}. The individual CCD images were cosmic ray cleaned when
possible using STSDAS tasks, which are packaged within IRAF
\footnote{IRAF is distributed by the National Optical
Astronomy Observatories, which is operated by the Association of
Universities for Research in Astronomy, Inc.\ under cooperative
agreement with the National Science foundation.}
.
These were then
combined using the STSDAS task {\it wmosaic}, which applies a field distortion correction
to each CCD image before combining them. An overlapping field (each lying within
a single CCD) was identified in both the first and second epoch images that 
contained both the star of suspected high velocity and a surrounding set of other
cluster stars.  In each case there were five nearby reference stars.
These cluster stars were used with IRAF task {\it geomap} to find the
relative orientation of the two images on the plane of the sky and then the two
were co-aligned using the task {\it geotran}, moving the first epoch image into 
alignment with the later image. The {\it geomap} task gives the one sigma values of
the accuracy of the relative position of the reference stars, which we use as the 
source of uncertainty values of our final measurements since the uncertainty of
measuring the position of the target stars was significantly smaller than these
values. In all cases the positions were determined using the IRAF task {\it imcntr}.
The position of the target star was then 
compared on the two images and from the difference in position the velocity in the plane of the sky was
calculated. For this calculation we assumed, as did PAH-A that the 
distance to the cluster is 470 pc. All of the images were in the low resolution
portion of the WFPC2, which means that motion across one pixel (0.0996\arcsec)
in one year would correspond to an angular motion of 9.96\arcsec /century or
222 \kms\ in the plane of the sky.

Since the reference stars have proper motions comparable to the JW average for the ONC,
the motions we derive will be relative to the cluster as a whole. G\'omez et~al. (2005) find from
radio astrometry that the spatial motion of the cluster is 5.4$\pm$0.6 \kms\ towards the 
south south-east and absolute motions for our stars would have to be corrected for this small value
of the cluster motion.

\subsection{JW 349} \label{mu349}

In the case of JW 349, the vector sum of the errors along the two axes from the 
{\it geomap} fit was 0.20 pixels and the measured difference of position of the
star was 0.17 pixels in the direction of Position Angle (PA=223\arcdeg). This
means that the formal value of the motion is less than the probable error of
measurement. The difference of epoch of the observations was 7.01 years, so that
we can conclude that the motion of JW 349 is no more than 0.28\arcsec/century or
6.3 \kms.

\subsection{JW 355} \label{mu355}

The vector sum of the errors of the {\it geomap} fit for JW 355 reference stars was
0.36 pixels and the measured difference of the star's position was 0.16 pixels towards 
PA=190\arcdeg. The difference of time in the observations was 10.06 years, so that 
we conclude that the motion of JW 355 is no more than 0.36\arcsec/century or
7.9 \kms.

\subsection{JW 451} \label{mu451}
The vector sum of the two axes errors of the fit was 0.20 pixels and the 
measured difference of the star's position was 0.28 pixels towards
PA=233\arcdeg. Since the difference in time of the observaions was 11.28 years, 
this means that the measured motion was only slightly greater
than the errors in its determination (3.9 \kms) and corresponds to 0.25\arcsec/century or 
5.5 \kms.

\section{Summary and Conclusions}
The motions given in \S\ 3 are in marked contrast with those reported in JW.
That publication gave vector motions of 1.72\arcsec/century
(38 \kms) for JW 349, 4.01\arcsec/century (89 \kms) for JW 355, and 3.08\arcsec/century (69 \kms) for JW 451.  The probable errors given in JW for these 
three stars are much larger than almost all the stars in their study, with 
vector summed errors of 0.38\arcsec/century, 0.41\arcsec/century, and 1.71\arcsec/century 
respectively. The differences between this HST study and JW are
large and outside the assigned errors of both studies. 

The cause of these 
differences is unknown, however, the background nebular emission is bright in
the region of each and the lower angular resolution of the groundbased material
could have made it difficult to measure the stars' positions. It may be that 
a contributor to the differences is that the three stars are all bright proplyds,
meaning that each has an extended ionized region surrounding it, thus making them
more difficult to measure at groundbased resolution. We note that 65 similar proplyds in the
JW study have a vector uncertainty of 0.83\arcsec/century, whereas the 648 JW stars that are not
known proplyds and have a probability of membership in the cluster of more than 90\%\ have a vector
uncertainty of only 0.16\arcsec/century. This summary of results from the JW study
indicates that there is a systematically larger uncertainty of position of the proplyds and
this may contribute to the large errors in the three stars of this study.

Another way of 
addressing the situation is to predict the changes on the HST images from the 
groundbased determined values. In this case, the expected motions would have been
1.2 pixels (JW 349), 4.3 pixels (JW 355), and 3.5 pixels (JW 451). Motions of
the size of the latter two predicted shifts would have been easily visible when comparing
the first and second epoch WFPC2 images and the 1.2 pixels predicted for JW 349
is much larger than the uncertainties in the HST measurement (0.2 pixels). 
On the basis of the quality of the WFPC2 images, we conclude that there is no
evidence that any of these three stars have a high velocity in the plane of the
sky.

We are forced, therefore, to conclude that the discussion of the status of JW 349,
JW 355, and JW 455 as low mass runaway stars as presented in PAH-A is not
valid, since the observations upon which they are founded are not supported
by our new HST results. 

G\'omez et~al (2005) find large motions for two of their radio sources near the ONC center.
Source 7 [GMR 14 \citep{gar87}, proplyd 155-338 \citep{ode96}]
and Source 14 [Zappata 46 \citep{zap04}, JW 503, and proplyd 160-353 \citep{ode96}]
have values of 16.5$\pm$2.4 \kms\ and 23.0$\pm$4.9 \kms\ respectively.
They question the former's high velocity on the basis that it could be due to changes in
the structure of this extended source and the latter's velocity because of possible
confusion with the signal from a nearby separate source. This leaves evidence for high 
velocity motion for only the sources near the BN-KL region.

Although this study disputes the measurements for the three optical candidate runaway
stars, search for others, using the WFPC2 would be worthwhile. This would require 
duplicating the pointing of the first epoch images, so that good astrometry could
be performed over entire CCD fields, rather than the fractional fields employed in
this investigation.


\acknowledgments
We wish to acknowledge that partial support for this work came from the Space 
Telescope Science Institute program GO 10246. That program and the timely 
production of its large amount of data would not have been possible without the 
special effort of the Orion Treasury Team at the Space Telescope Science Institute,
in particular David Soderblom and Eddie
Bergeron. We are grateful to Luis Rodr\'{\i}guez for suggesting that we examine
the WFPC2 images for runaway objects. We particularly appreciate the efforts of
Lick Observatory's Burton Jones for reassessing the errors in his groundbased study
of this region and his encouragement to procede with publishing 
these new results.

{\it Facilities:} \facility{HST (WFPC2)}.





\begin{thebibliography}{}
\bibitem[Bally et~al.(2000)]{bal00} Bally, J., O'Dell, C. R., \& McCaughrean, M. J. 2000, \aj, 119, 2919
\bibitem[Holtzman et~al.(1995)]{hol95} Holtzman, J. A., Burrows, C. J., Casertano. S., Hester, J. J., Trauger, J. T., Watson, A. M., \& Worthey, G. 1995, PASP, 107, 1065
\bibitem[Garay, Moran, \& Reid (1987)]{gar87} Garay, G., Moran, J. M., \& Reid, M. J. 1987, \apj, 314, 535
\bibitem[G\'omez et~al.(2005)]{gom05} G\'omez, L., Rodr\'{\i}guez, L. F., Loinard, L, Lizano, S., Poveda, A., \& Allen, C. 2005, \apj, in press
\bibitem[Jones \& Walker(1988)]{jon88} Jones, B. F. \& Walker, M. F. 1988,\aj, 95, 1755 (JW)
\bibitem[O'Dell et~al.(2002)]{ode02} O'Dell, C. R., Doi, T. \& Hartigan, P. 2002, \aj, 124, 445
\bibitem[O'Dell et~al.(2003)]{ode03} O'Dell, C. R., Peimbert, M., \& Peimbert, A. 2003, \aj, 125, 2590
\bibitem[O'Dell \& Wen(1994)]{ode94} O'Dell, C. R., \& Wen, Z. 1994, ApJ, 436, 194
\bibitem[O'Dell \& Wong(1996)]{ode96} O'Dell, C. R., \& Wong, S.-K. 1996, \aj, 111, 846
\bibitem[Poveda et~al.(2005)]{pov05} Poveda, A., Allen, C., \& Hern\'andez-Alc\'antara, A.
    2005, \apj, 627, L61 (PAH-A)
\bibitem[Poveda et~al.(1967)]{pov67} Poveda, A., Ruiz, J., \& Allen, C. 1967, Bol. Obs. Tonantzintla Tacubaya, 4, 86, http://www.astrosmo.unam.mx/~luisfr/Poveda67.pdf
\bibitem[Robberto et~al. (2004)]{rob04} Robberto, M., Soderblom, D. R., O'Dell, C. R., Stassun, K. G., Hillenbrand, L. A., Simon, M., Feigelson, E. D., Najita, J., and 13 coauthors 2004, AAS, 205, 117.3
\bibitem[Rodr\'{\i}guez et~al.(2005)]{rod05} Rodr\'{\i}guez, L. F., Poveda, A., Lizano, S., \& Allen, C. 2005, \apj, 627, L65
\bibitem[Tan(2004)]{tan04} Tan, J. C. 2004, ApJ, 607, L47
\bibitem[Zappata et~al. (2004)]{zap04} Zappata, L. A., Rodr\'{\i}guez, L. F., Kurtz, S. E., \& O'Dell, C. R. 2004, \aj, 127, 2252
\end{thebibliography}
\end{document}